\documentstyle[12pt,aasms4] {article}
\newcommand{\beq}{\begin{equation}}
\newcommand{\eeq}{\end{equation}}
\newcommand{\beqn}{\begin{eqnarray}}
\newcommand{\eeqn}{\end{eqnarray}}

\begin{document}
\title{\bf{The Drag Induced Resonant Capture for Kuiper Belt Objects} }
\author{Ing-Guey Jiang$^{1}$ and Li-Chin Yeh$^{2}$}

\affil{
{$^{1}$ Institute of Astronomy,}\\
{ National Central University, Chung-Li, Taiwan}\\
{$^{2}$ Department of Mathematics,}\\
{ National Hsinchu Teachers College, Hsin-Chu, Taiwan} 
}

\authoremail{jiang@astro.ncu.edu.tw}

\begin{abstract}
It has been an interesting question that 
why there are one-third of Kuiper Belt Objects (KBOs) trapped into the 
3:2 resonance but, in contrast,  
only several KBOs are claimed to be associated with the 2:1 resonance.
In a model proposed by Zhou et al. (2002), 
the stochastic outward migration of the Neptune
could reduce the number of particles in the 2:1 resonance
and thus the 
objects in the 3:2 resonance become more distinct.  
As a complementary study, we investigate  
the effect of proto-stellar discs on the resonance capture.
Our results show that
the gaseous 
drag of a proto-stellar disc can trap KBOs into 
the 3:2 resonance rather easily. 
In addition, no objects are captured into the 2:1 resonance in our simulation.

\end{abstract}


\newpage
\section{Introduction}

The general picture of the outer solar system has been changed due to the
discovery of KBOs 
(Jewitt and Luu 1993, Williams et al. 1995). 
The origin of the dynamical properties of 
KBOs has become a very interesting but controversial issue.
Particularly, it is surprising that one-third of the population
are engaged into the 3:2 resonance with the Neptune.  
 
Malhotra (1995) proposed a mechanism of resonance sweeping to explain 
the large fraction of 3:2 resonance objects, plutinos, in the Kuiper Belt.
In that model, as the Neptune radially migrate outward, its mean 
motion resonance swept ahead of it through the Kuiper Belt and captured the 
Pluto, along with plutinos, into the 3:2 resonance.
This theory predicted that the populations of the 3:2 and 2:1 resonance
should be of the same order. Moreover, the radial migration
of the Neptune is assumed to be smooth and continuous.  
Zhou et al. (2002) improved this by introducing 
a stochastic term to make this outward migration 
more realistic. They showed that for particular chosen parameters, the
number of particles in the 2:1 resonance can be reduced and thus the 
objects in the 3:2 resonance become more distinct.  

Yeh and Jiang (2001) showed that the scattered planet shall move on an 
eccentric orbit and thus the pure radial migration is too naive, although it 
could be an approximation. 
For example,  
the Neptune's orbit is not circular during its outward migration
as shown in the simulations by Thommes et al. (1999).
However, 
the Neptune's current orbit is nearly circular and thus one
needs to introduce a massive Kuiper Belt to 
circularize the Neptune's orbit. Please also see the simulations
in Gomes (2003).


In fact, 
the total mass of Kuiper Belt is itself a controversial issue. In order
to form  100 km size KBOs around the current region, one would 
need a much more massive Kuiper Belt initially
because the planetesimal accretion rate has to be high enough
to make KBOs formed in time.
However, the material in the Kuiper Belt has to be depleted 
significantly during the evolution 
because the observational upper limit of the mass at the Kuiper 
Belt is only 0.1 Earth mass.  
To avoid this formation problem, Levison and 
Morbidelli (2003) 
proposed a pushing-out model that both the Kuiper Belt Objects
and Neptune were forced to migrate outward from where they were
formed initially. They showed that some 
KBOs were captured into
2:1 resonance and could 
move out together with the Neptune in their simulations.

On the other hand, the traditional planet formation theory, 
i.e. core-accretion model, is facing the
competition from the disc-instability model (Boss 1998). 
Because the disk-instability model can form a Jupiter
mass planet in a few thousand years, it could be that the formation of 
the Neptune and KBOs is not that difficult as the traditional model. Although
the solid core formation through dust grain growth probably still play an
important role, the condition to initiate the formation of Neptune and KBOs
and thus the formation timescales could be very different. 
Furthermore, Bryden et al. (2000) showed that even by the core-accretion
model, it is possible to form the Neptune around the current location 
in $10^6$ or $10^7$ years, which is 
comparable or shorter than the disc depletion timescale. 
In any case, the higher density gaseous blobs might
shorten the dust-accretion process. Therefore, it is not clear 
how severe the initial mass problem is.  

No matter what the exact formation scenario is,
our model here could represent a process that starts at the time (a) when the 
proto-Neptune and proto-KBOs were already pushed out after a violent migration 
and settle down around current places (if it is  difficult 
to form at current locations)  or (b) after the Neptune and KBOs
have grown and formed around current regions
with only very gentle migrations.

We are particularly interested in the influence from the proto-stellar
disc and thus the possible migration is ignored here.
The model will be described briefly in Section 2 and 
the results would be in Section 3. There are some discussions in 
Section 4.
Section 5 concludes the paper.

\section{The Model}

In our model, the test particle moves in a disc-star-planet system: 
both the star and planet are doing circular motions with fixed 
star-planet distance  and a disc is added to provide additional 
forces on the test particle.
Since we consider the two dimensional coplanar orbits only, 
the equations of motion are similar to the ones used in 
Jiang and Yeh (2004a)
but only that we here use a power-law density profile for the disc as in 
Jiang and Yeh (2004b). 

Moreover, we also introduce the drag force as in 
Murray and Dermott (1999).  We choose a formula
that the drag force is proportional to the local disc density and 
also the difference between  
the test particle's
velocity and the disc's local rotating velocity (assuming 
a Keplerian disc). That is, for a test particle located at $(\xi,\eta)$
with velocity $(d\xi/dt, d\eta/dt)$
in the inertial frame, 
the drag force per unit mass is
\beq
{\bf F}=-\alpha \Sigma(r) (\frac{d\xi}{dt}-v_{\xi}){\hat \xi}
-\alpha \Sigma(r) (\frac{d\eta}{dt}-v_{\eta}) {\hat \eta},
\eeq
where $(v_{\xi}, v_{\eta})$ is the disc's local rotating velocity and
$\Sigma(r)$ is the disc surface density, which is a function of 
radius only.
The proportional constant $\alpha$ is set to be 1/5.

The disc mass is assumed to be $M_b=0.01$.
900 test particles are randomly placed in a belt region 
$1.1 \le r \le 2.0$ with uniform number distribution, where $r$ is the
radial coordinate. They are all doing circular motions initially. 
The simulation starts from $t=0$ and stop 
at $t=t_{\rm end}=123200\pi\sim 3.8\times 10^5$, which is about $10^7$ years 
in the real time scale when the unit of mass is $M_{\odot}$ and 
the unit of length is 30 AU.

\section{The Results}

Figure 1 and Figure 2 are the results of our simulation. 
In Figure 1, crosses mark the initial locations of all test 
particles in the $x-y$ plane, i.e. orbital plane. 
We calculate the 3:2 resonant argument 
$\phi$ (and also the 2:1 resonant argument $\phi_1$)
for any $t$ during the simulation for all particles.
We determine both the maximum 
and the minimum of the 3:2 resonant argument $\phi$, 
i.e. $\phi_{\rm max}$ and $\phi_{\rm min}$,
respectively for the final stage when $t\in [t_{\rm f}, t_{\rm end}]$
(where $t_{\rm f}=3.5\times 10^5$).
If $\phi_{\rm max}-\phi_{\rm min}<180^o$, 
this particle is in 3:2 resonance for our
simulation. We then use
circles to mark the initial locations of test particles which
are in the 3:2 resonance. These circles overlap some of the crosses. 
The same procedure is done for the 2:1 resonance but
we find that no particle is captured into the 2:1 resonance.

To investigate more details,
Figure 2 provides a combination of different aspects of the results.
In Figure 2(a), we show the positions of all particles on the $a-e$ plane
when $t=t_{\rm end}$. It appears that many particles' orbital eccentricities
are amplified to a larger value for those with $a< 1.35$. 
Particularly, the particles
with $a=1.33$ have the largest eccentricities among these particles. 
There is also a small number
of particles with slightly larger eccentricities around $a=1.6$.
We find that, in fact, the region of the 3:2 resonance is close to $a=1.33$
and the region of the 2:1 resonance is near $a=1.6$. In addition to that,
there are also concentrations of particles in the regions with $a<1.33$.
These regions are corresponding to the 7:5, 4:3, 5:4 and 6:5 resonances
approximately. These particles could have been influenced by these resonances.
 
When we plot the positions of all test particles on the $x-y$ plane
at $t=t_{\rm end}$ as in Figure 2(b), we find that, in deed, 
there is a concentration of particles at a particular radius about $1.33$.
Thus, it is likely that 
many particles are captured into the 3:2 resonance. During our simulation,
all the particles are forced to migrate inward
due to the gas drag.  For the particles with initial $a>1.33$, they stop this
inward migration at $a=1.33$ due to the 3:2 resonance capture. 
That explains why there are many particles at $a=1.33$ with higher 
eccentricities at $t=t_{\rm end}$.
To understand more about this drag induced inward migration and
resonance capture, we make Figure 2(c), in which we plot the number of 
particles captured into the resonance as a function of time.  
The number of particles in a particular resonance at $t_i$ is defined to be 
the total number of particles with 
the difference between the maximum and minimum resonance arguments 
less than $180^o$
during $t_{i-1}< t < t_i$. We set  
$t_0=0, t_i-t_{i-1}= 11200\pi$, where $i=1,2,...,11$.
We did calculations for both 3:2 and 2:1 resonances. We find that 
the number of particles captured into the 
3:2 resonance keeps increasing and finally becomes
a very large fraction of all particles as shown in Figure 2(c). 
However, we still find that 
no particle gets captured into the 2:1 resonance during the simulation.
Figure 2(d) shows the 3:2 resonance argument as a function of time for a 
particle initially located at (x,y)=(0.940, 1.047) marked in Figure 2(b).
It shows that this particle gets captured into the 
3:2 resonance after $t=t_2$.\\

\section{Discussions}

Our results show that the effect of the drag force from the disc is important. 
As a comparison, let us discuss the situation when there is no disc.
In this case, the test particles are influenced by the central star
and the Neptune only. Most particles will continue to do circular motions,
although the secular perturbation from the Neptune 
might affect them but this process is very slow.
There are two kinds of particles which might change their orbits quickly:
(1) those are very close to the Neptune
(They are influenced by the Neptune quickly and might get scattered.);
(2) those in the resonant regions 
(Their eccentricity might get increased. Part of them might leave
and some of them might stay in the resonant regions.).
There would be no obvious inward migrations for test particles. Thus, 
the probability for them to get captured into 3:2 resonance would 
be much smaller.

The speed of inward migration shall be proportional to 
the strength of the drag force.
Thus, the stronger force could make more particles migrate and get captured
by the 3:2 resonance given that 
there are enough particles in the outer part of the Kuiper Belt. 
Therefore, the results could be related to the  
particles' initial distribution and also the strength of the drag force.

Because the proto-stellar disc definitely exists and provides the drag force
when the proto-KBOs are forming, the mechanism of drag induced resonant
capture seems to be attractive and cannot be ignored.
It explains the resonant KBOs in a natural way.
The traditional mechanism of sweeping capture 
by the migrating Neptune also works well as 
demonstrated by 
the models of stochastic migrations in Zhou et al. (2002). 
However, in that case, one would need a physical mechanism to explain
the Neptune's outward migration. One also has to understand how
the migration stops, thus explains the Neptune's current orbital radius and 
how the orbit could become nearly circular finally.

The details of the new mechanism of drag induced resonant capture
and its combination with the traditional sweeping capture by the 
migrating Neptune
would be an interesting future work.


\section{Concluding Remarks}

We have used a model of the disc-star-planet system
to investigate the effect of a gaseous proto-stellar disc on the resonance 
capture. In addition to the force from the central star and planet, the test 
particle is also influenced by the gravitational and frictional forces
from the disc.
Particularly, we apply our model to the problem of resonant KBOs and focus
on the 3:2 and 2:1 resonances.

Our results show that the drag force plays an important role for 
the resonant capture and many KBOs can get captured into the 
3:2 resonance but not into the
2:1 resonance. 
This is consistent with the observation that about 
one-third of the whole population
of KBOs are in the 
3:2 resonance and only a few are claimed to be in the 2:1 resonance.
Therefore, the mechanism we study here might be helpful to understand 
the history of KBOs' resonance capture.

\section*{Acknowledgment}

We thank James Binney's participation in the 2004 Annual Meeting
of Chinese Astronomical Society, Taipei, where we presented part of our
results and gained good comments and suggestions from him for our group's
projects. This work is supported in part 
by National Science Council, Taiwan, under 
Ing-Guey Jiang's Grants: NSC 92-2112-M-008-043 and also under Li-Chin Yeh's 
Grants: NSC 92-2115-M-134-001.
The numerical simulations are done by the PC Cluster, IanCu,
located in Institute of Astronomy, National Central University.

\clearpage

\begin{figure}[tbhp]
\epsfysize 6.5 in \epsffile{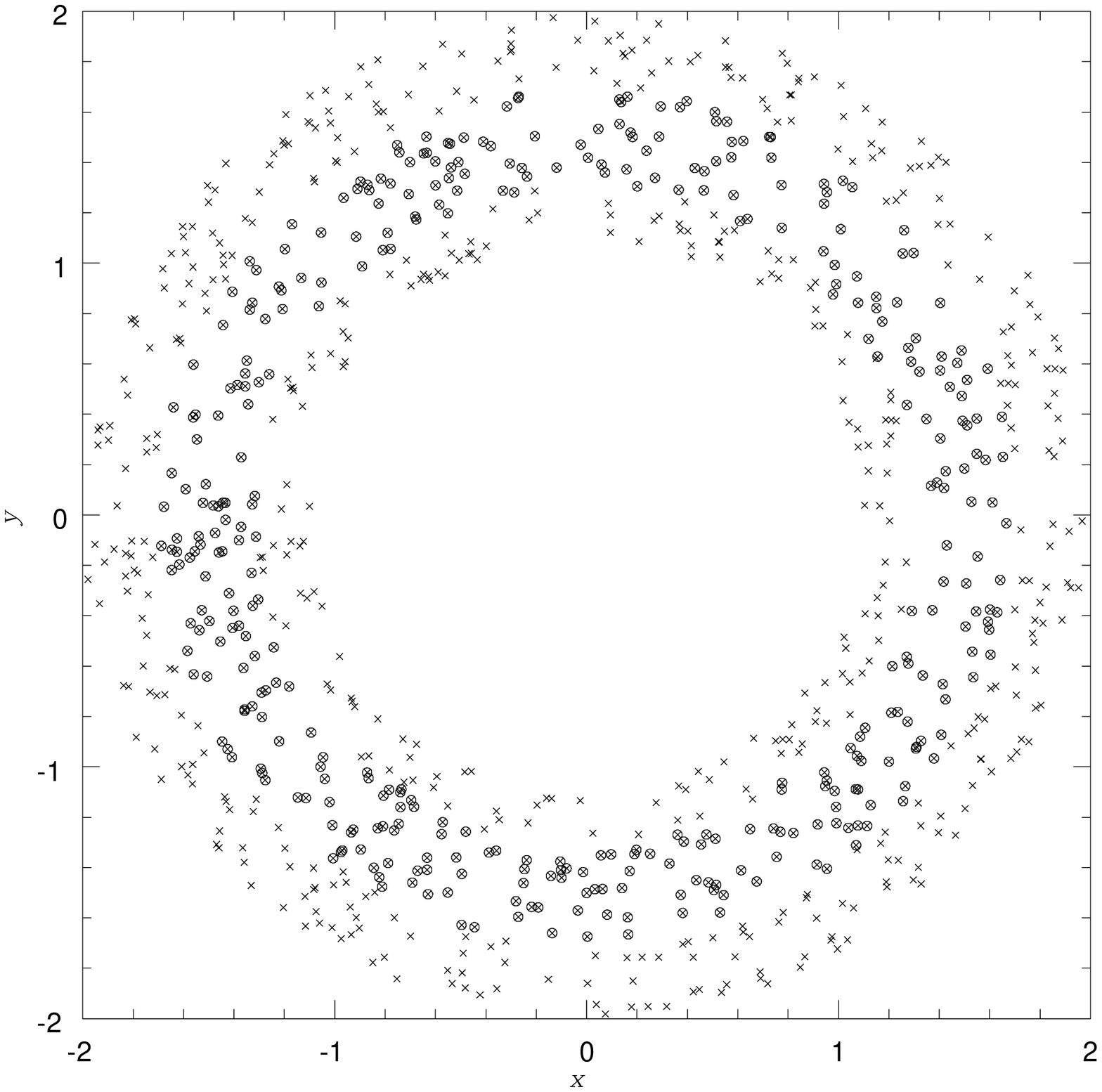}
\caption{The initial locations of all test particles (crosses) 
and also the initial locations of those particles which are captured into 3:2
resonance (circles).
}
\end{figure}

\begin{figure}[tbhp]
\epsfysize 6.5 in \epsffile{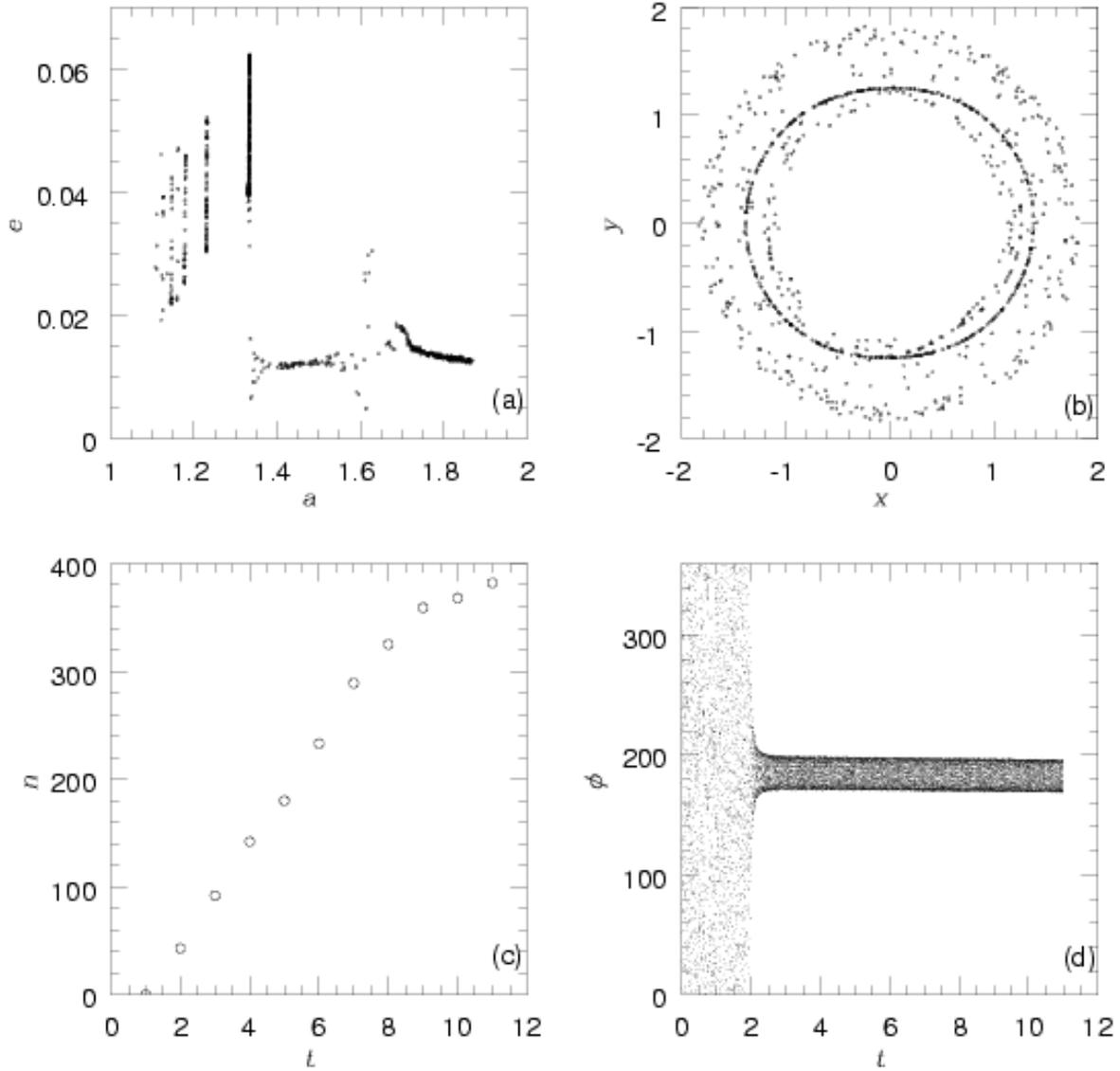}
\caption{(a) The test particles on the $a-e$ plane
when $t=t_{\rm end}$. (b) The test particles on the $x-y$ plane
when $t=t_{\rm end}$. (c) The number of test particles captured
in the 3:2 resonance as a function of time. (d) The 3:2 resonance argument
as a function of time for a particular test particle. (Please see
the main text for the details of (c) and (d) and their unit of time is 
$11200\pi$.)
}
\end{figure}

\end{document}